\documentclass[12pt, journal, letterpaper, onecolumn, draftclsnofoot]{IEEEtran}

\IEEEoverridecommandlockouts

\usepackage{graphicx}
\usepackage{subfigure}
\usepackage{curves}
\usepackage{amsfonts}
\usepackage[cmex10]{amsmath}
\usepackage{epsfig}
\usepackage{stfloats}
\usepackage{amssymb}
\usepackage{cite}
\usepackage{epstopdf}
\usepackage{bm}
\usepackage{bbm}
\usepackage{algorithm}
\usepackage{algpseudocode}

\begin{document}

\title{Hybrid Polar Encoding with Applications in Non-Coherent Channels}
\author{\IEEEauthorblockN{Mengfan~Zheng}\\
	\IEEEauthorblockA{Department of Electronic and Computer Engineering\\The Hong Kong University of Science and Technology\\Email: eemzheng@ust.hk}
}

\maketitle

\begin{abstract}
	In coding theory, an error-correcting code can be encoded either systematically or non-systematically. In a systematic encode, the input data is embedded in the encoded output. Conversely, in a non-systematic code, the output does not contain the input symbols. In this paper, we propose a hybrid encoding scheme for polar codes, in which some data bits are systematically encoded while the rest are non-systematically encoded. Based on the proposed scheme, we design a joint channel estimation and data decoding scheme. We use the systematic bits in the hybrid encoding scheme as pilots for channel estimation. To mitigate the code rate loss caused by the pilots and to provide additional error detecting capability, we propose a dynamic pilot design by building connections between the systematic bits and non-systematic bits. Simulation results show that the performance of the proposed scheme approaches that of the traditional non-systematic polar coding scheme with perfect channel state information (CSI) with the increase of SNR.
\end{abstract}

\section{Introduction}

Polar codes \cite{arikan2009channel} are the first family of capacity achieving channel codes and have been adopted in the fifth generation wireless communication standards. One of the key differences of polar codes from traditional channel codes is their channel-specific design, indicating that the performance of polar codes heavily relies on the knowledge of channel state information (CSI). Real-world wireless communication channels are usually time-varying, and polar codes designed for static channels may not be able to achieve the desired performance.  

The fading channel is a classic model for time-varying channels. In the fading channel, the channel gain changes over time according to some distribution, known as the channel distribution information (CDI). The block fading channel models the case that the channel gain remains constant over a fixed time interval $T_c$, called channel coherent time, and changes to another independent value afterwards. 
Traditionally, the CSI is first estimated by sending some known symbols, called pilots. Then data is encoded and transmitted based on the estimated CSI. Such an approach is close to optimal when the channel coherent time is very long. However, when the channel coherent time is small, inserting pilots in the transmitted signal significantly lowers the overall data rate. In this case, non-coherent communication can be a better choice.

There have been some researches on polar coding for fading channels without CSI in the literature. Polar coding for i.i.d. fading channels with only CDI was considered in \cite{Bravo2012fading}. A multi-level polar coding scheme for non-coherent block fading channels is proposed in \cite{zheng2018NonC}, which is only suitable for block fading channels with very small coherent time. Reference \cite{Li2018CESys} proposed a channel estimation scheme based on systematic polar codes by selecting pilot symbols from the coded symbols. A list decoding scheme that can estimate the channel coefficient and decode message simultaneously with the help of a few pilots was proposed in \cite{Xhemrishi2019LDSCC}, which requires a large list size to achieve good performance. A pilot-free two-stage polar coding scheme that jointly estimates the CSI and data based on the constraints imposed by the frozen bits was proposed in \cite{Yuan2021PCNonC}, which can approach the performance of a coherent receiver with perfect CSI.

All the aforementioned schemes are based on traditional non-systematic or systematic polar codes. In this paper, we design a new polar coding scheme for non-coherent communications over block fading channels based on a hybrid encoding construction that combines non-systematic and systematic polar encoding. In this construction, part of the message bits are encoded non-systematically and the rest are encoded systematically. The idea of such a hybrid polar encoding design has been proposed in \cite{jin2018partial} for the purpose of reducing encoding and decoding resource consumption (called partial systematic polar codes in that paper). In our proposed scheme, the systematically encoded message bits serve as dynamic pilots which will be decoded first and used for channel estimation. Then the rest message bits are decoded with the estimated CSI. By exploiting the connection between the systematic and non-systematic bits and with a properly designed two-phase decoder, we show by simulations that the proposed scheme can approach the performance of the traditional non-systematic polar coding scheme with perfect CSI as the SNR increases and can even outperform it at high SNRs. We discuss the reason for this phenomenon in details with our simulation results.

\textit{Notations:} $[N]$ is the abbreviation for $\{1,2,...,N\}$. Vectors and matrices are denoted by lowercase and uppercase boldface letters, respectively. Vectors can also be denoted as $x^{a:b}\triangleq [x_a,x_{a+1},...,x_{b}]$ for $a\leq b$. For $\mathcal{A}\subset [N]$ and $\mathcal{B}\subset [N]$, $x^{\mathcal{A}}$ denotes the subvector $[x_i:i\in\mathcal{A}]$ of $x^{1:N}$, $\mathbf{G}^{\mathcal{A}}$ is the submatrix of $\mathbf{G}$ consisting of rows with indices in $\mathcal{A}$, and $\mathbf{G}^{\mathcal{A}\mathcal{B}}$ is the submatrix of $\mathbf{G}$ consisting of elements $G_{ij}$, where $i\in\mathcal{A}$ and $j\in \mathcal{B}$.

\section{Preliminaries}
\label{Sec:Preli}

\subsection{System Model}

We consider communications over a block fading channel with coherent time $T_c$. Channel coding is performed across $B$ coherent blocks. The frame size is then $N=BT_c$. At block $i$ ($i=1,...,B$), the channel is modelled as
\begin{equation}
	\mathbf{y}^{(i)}=h_i\mathbf{x}^{(i)}+\mathbf{w}^{(i)},
\end{equation}
where $h_i\in \mathbb{R}$ is the channel gain at block $i$, $\mathbf{x}^{(i)}=[x^{(i)}_1,x^{(i)}_2,...,x^{(i)}_{T_c}]$ is the channel input, $\mathbf{y}^{(i)}=[y^{(i)}_1,y^{(i)}_2,...,y^{(i)}_{T_c}]$ is the channel output, and $\mathbf{w}^{(i)}=[w^{(i)}_1,w^{(i)}_2,...,w^{(i)}_{T_c}]$ is the white Gaussian noise, with $w^{(i)}_j\sim \mathcal{N}(0,\sigma^2)$ for $j\in[T_c]$. We assume BPSK modulation is used in this paper, i.e., $x^{(i)}_j\in\{-1,1\}$, although the proposed scheme can be easily extended to higher order modulation scenarios. 
We study the non-coherent case when neither the transmitter nor the receiver has the instantaneous CSI of the channel. The CDI is assumed to be known by the receiver.

\subsection{Polar Codes}

Polar codes are defined by the following polar transform \cite{arikan2009channel}
\begin{equation}
	\mathbf{x}=\mathbf{u}\mathbf{G}_N, \label{encoding}
\end{equation}
where $\mathbf{u}=[u_1,...,u_N]$ is the uncoded bit sequence, $\mathbf{x}=[x_1,...,x_N]$ is the encoded codeword and $\mathbf{G}_N=\textbf{F}^{\otimes n}$. Here $\textbf{F}=
\begin{bmatrix}
	1 & 0 \\
	1 & 1
\end{bmatrix}$ and $\otimes$ is the Kronecker power. $\mathbf{G}_N$ can also be written as $\mathbf{B}_N \textbf{F}^{\otimes n}$ with $\mathbf{B}_N$ being the bit-reversal matrix. However, for systematic encoding it is more convenient to use the generator matrix without $\mathbf{B}_N$.

\subsubsection{Non-Systematic Encoding}
The non-systematic construction of polar codes partitions $\mathbf{u}$ into an \textit{information set} $\mathcal{I}$ and a \textit{frozen set} $\mathcal{F}=\mathcal{I}^c$, where $\mathcal{I}^c$ denotes the complementary set of $\mathcal{I}$. Message bits are assigned to $\mathbf{u}^{\mathcal{I}}$ while $\mathbf{u}^{\mathcal{F}}$ are assigned with some fixed value, such as 0. Then the encoding process of (\ref{encoding}) can be written as
\begin{equation}
	\mathbf{x}=u^{\mathcal{I}}\mathbf{G}_N^{\mathcal{I}}+u^{\mathcal{F}}\mathbf{G}_N^{\mathcal{F}}. \label{PolarEnc}
\end{equation}

\subsubsection{Systematic Encoding}
In systematic encoding, message bits directly appear in $\mathbf{x}$. Suppose $\mathbf{x}^{\mathcal{J}}$ ($\mathcal{J}\subset [N]$) are used to carry message bits, where $|\mathcal{J}|=|\mathcal{I}|$. It is shown that if $\mathbf{G}_N^{\mathcal{I}\mathcal{J}}$ is invertible \cite{Arikan2011sys}, then
\begin{equation}
	u^{\mathcal{I}}=(x^{\mathcal{J}}-u^{\mathcal{I}^c}\mathbf{G}_N^{\mathcal{I}^c\mathcal{J}})(\mathbf{G}_N^{\mathcal{I}\mathcal{J}})^{-1}. \label{SysEnc}
\end{equation}
Thus, to perform systematic polar encoding, one can first assign $u^{\mathcal{I}^c}$ with the frozen value, compute $u^{\mathcal{I}}$ according to (\ref{SysEnc}) and then obtain $\mathbf{x}$ according to (\ref{encoding}). Note that the encoding complexity of this approach is higher than that of non-systematic encoding due to the matrix inversion operation. For low-complexity systematic polar encoding algorithms, we refer the readers to \cite{Sarkis2016Sys,Vangala2016Sys,Chen2016Sys}.

\subsubsection{Decoding}
Upon receiving $\mathbf{y}$, the receiver can use a successive cancellation (SC) decoder to recover $\mathbf{u}$:
\begin{equation}
	\hat{u}_{i}=
	\begin{cases}
		u_i,&\text{ if } i\in \mathcal{F}\\
		\arg\max_{u\in\{0,1\}}P_{U_{i}|\mathbf{Y}, U^{1:{i-1}}}(u|\mathbf{y},\hat{u}^{1:{i-1}}),&
		\text{ if } i\in \mathcal{I}
	\end{cases}.
\end{equation}
For non-systematic encoding, the original information is retrieved from $\hat{u}^{1:N}$. For systematic encoding, the original information is retrieved from $\hat{x}^{1:N}=\hat{u}^{1:N}\mathbf{G}_N$.

To improve error performance, we can use the cyclic redundancy check (CRC)-aided successive cancellation list (CA-SCL) decoding \cite{Niu2012CA,tal2015list}. The idea of CA-SCL is to retain up to $L$ most probable paths during the SC decoding process and use CRC to eliminate wrong paths.

\subsection{Hybrid Polar Encoding}
\label{Sec:HybScheme}
In the hybrid polar encoding scheme, part of the message bits are mapped to $\mathbf{u}$, while the rest are directly shown in $\mathbf{x}$.
The hybrid polar encoding goes as follows. Denote the non-systematic bit positions in $\mathbf{u}$ by $\mathcal{I}_{ns}$ and the systematic bit positions in $\mathbf{x}$ by $\mathcal{I}_{s}$. We assign frozen bits $\mathbf{u}^{\mathcal{F}}$ to 0. Then we have
\begin{align}
	x^{\mathcal{I}_{s}}&=u^{\mathcal{I}_{ns}}\mathbf{G}_N^{\mathcal{I}_{ns}\mathcal{I}_{s}}+u^{\mathcal{I}_{s}}\mathbf{G}_N^{\mathcal{I}_{s}\mathcal{I}_{s}},\label{Hyb-1}\\
	x^{\mathcal{I}_{ns}}&=u^{\mathcal{I}_{ns}}\mathbf{G}_N^{\mathcal{I}_{ns}\mathcal{I}_{ns}}+u^{\mathcal{I}_{s}}\mathbf{G}_N^{\mathcal{I}_{s}\mathcal{I}_{ns}},\label{Hyb-2}\\
	x^{\mathcal{F}}&=u^{\mathcal{I}_{ns}}\mathbf{G}_N^{\mathcal{I}_{ns}\mathcal{F}}+u^{\mathcal{I}_{s}}\mathbf{G}_N^{\mathcal{I}_{s}\mathcal{F}}.\label{Hyb-3}
\end{align}
$x^{\mathcal{I}_{s}}$ and $u^{\mathcal{I}_{ns}}$ are used to carry information. From (\ref{Hyb-1}) we have
\begin{equation}
	u^{\mathcal{I}_{s}}=(x^{\mathcal{I}_{s}}-u^{\mathcal{I}_{ns}}\mathbf{G}_N^{\mathcal{I}_{ns}\mathcal{I}_{s}})(\mathbf{G}_N^{\mathcal{I}_{s}\mathcal{I}_{s}})^{-1}.\label{Hyb-4}
\end{equation}
Then $x^{\mathcal{I}_{ns}}$ and $x^{\mathcal{F}}$ can be obtained according to (\ref{Hyb-4}), (\ref{Hyb-2}) and (\ref{Hyb-3}).

\section{Joint Channel Estimation and Decoding with Hybrid Polar Codes}
\begin{figure}[htb]
	\centering
	\includegraphics[width=0.6\columnwidth]{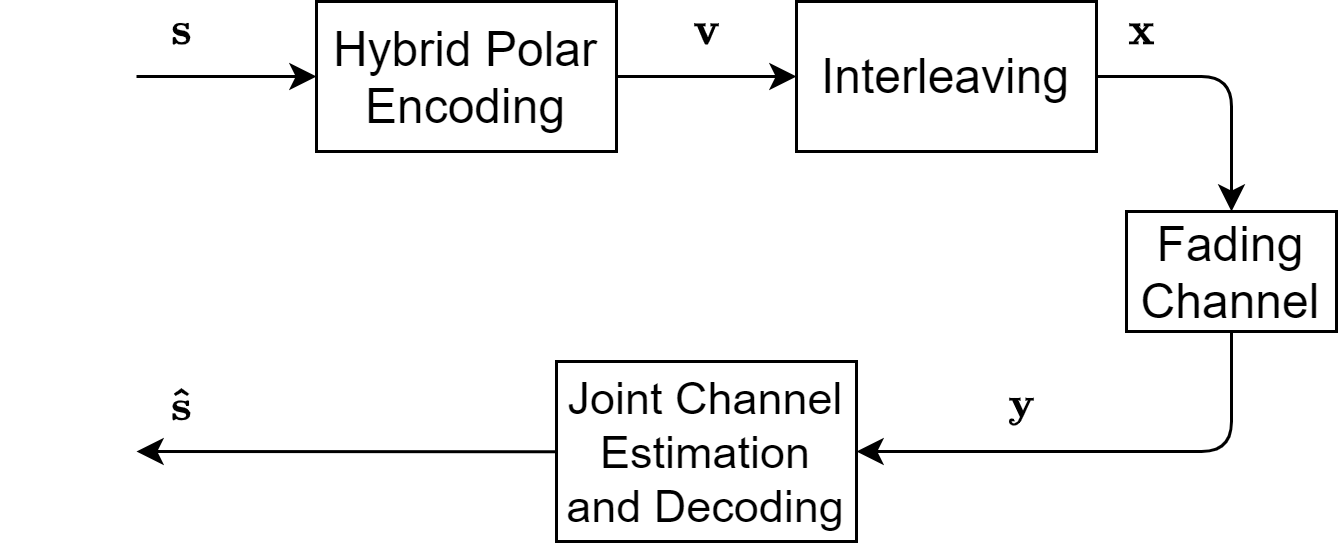} 
	\caption{The proposed scheme.} 
	\label{fig:scheme}
\end{figure}
\begin{figure}[htb]
	\centering
	\includegraphics[width=0.6\columnwidth]{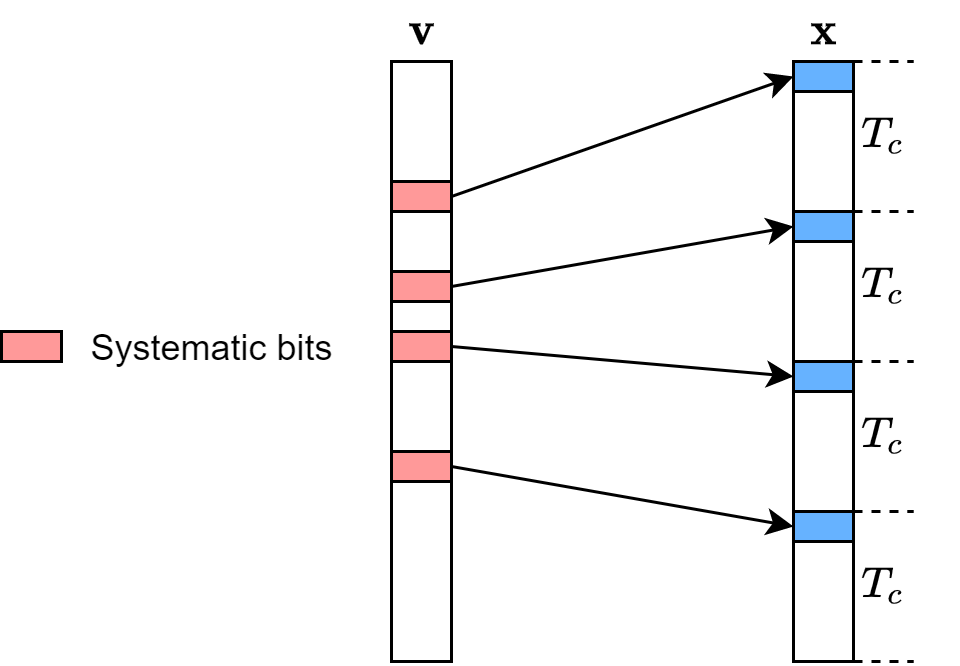} 
	\caption{Interleaver design.} 
	\label{fig:mapping}
\end{figure}

Our proposed scheme is illustrated in Fig. \ref{fig:scheme}. 
$\mathbf{s}$ is the message vector to be sent and $\mathbf{v}$ is the output vector of the hybrid polar encoder. The transmitted sequence in a frame is denoted by $\mathbf{x}\triangleq [\mathbf{x}^{(1)},...,\mathbf{x}^{(B)}]$, which is an interleaved version of $\mathbf{v}$. The received signal is denoted by $\mathbf{y}\triangleq [\mathbf{y}^{(1)},...,\mathbf{y}^{(B)}]$. Let $\mathbf{u}=\mathbf{v}\mathbf{G}_N$ and denote the non-systematic bit set, systematic bit set and frozen bit set by $\mathcal{I}_{ns}$, $\mathcal{I}_{s}$ and $\mathcal{F}$, respectively. Suppose the code rate is $R$. Then the number of message bits in each frame is $K=NR=BT_cR$. In our scheme, we use the systematic bits to perform channel estimation. Thus, we also refer to them as pilots in the sequel. The purpose of the interleaver is to distribute the pilots uniformly among the coherent blocks. An example is shown in Fig. \ref{fig:mapping}, in which the pilots are mapped to the beginning of each coherent block. 

A straightforward approach to apply hybrid polar coding for non-coherent communications is to assign the systematic bits with some known value. The decoder uses them to perform channel estimation and then recovers the message embedded in the non-systematic bits with the estimated channel gains. The drawback of this approach is obvious. Since the systematic bits does not carry message, to maintain a target message transmission rate one has to increase the rate of the polar code, which will deteriorate error performance. 

\subsection{Dynamic Pilot Design}

\begin{figure}[tb]
	\centering
	\includegraphics[width=0.6\columnwidth]{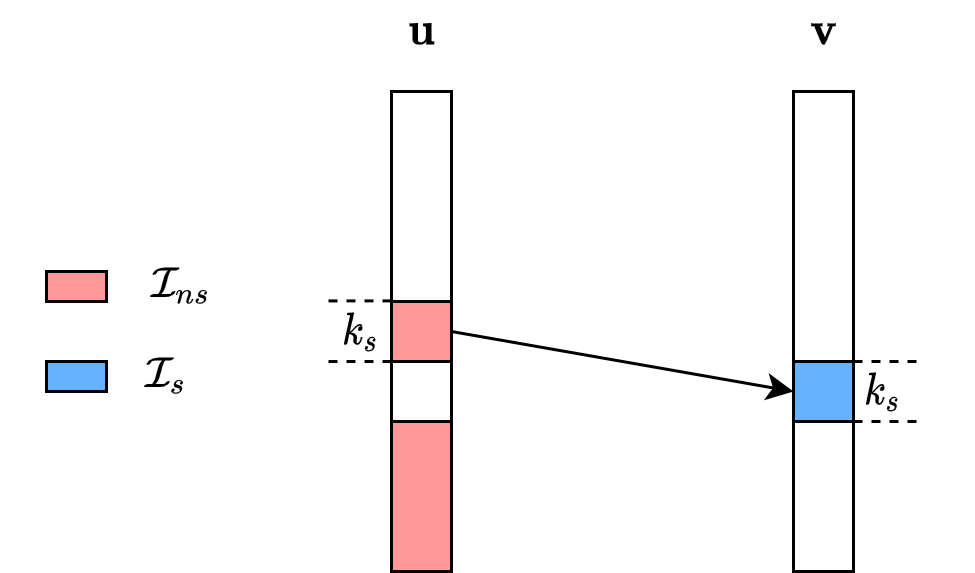} 
	\caption{Dynamic pilot design.} 
	\label{fig:hybrid}
\end{figure}

To utilise the systematic bits as pilots while not sacrificing the error correcting capability, we propose the following dynamic pilot design. Let $k_s=|\mathcal{I}_{s}|$ and denote $\mathcal{I}=\mathcal{I}_{ns}\cup \mathcal{I}_s$. In the proposed scheme, we require that the first $k_s$ indices in $\mathcal{I}$, denoted by $\mathcal{I}^{(1:k_s)}$, all belong to $\mathcal{I}_{ns}$. This is to ensure that in an SCL decoder, the first $k_s$ non-frozen bits to be decoded (i.e., $u^{\mathcal{I}^{(1:k_s)}}$) all belong to the non-systematic bits. At the encoder side, a message vector $\mathbf{s}$ together with its CRC bits are assigned to $u^{\mathcal{I}_{ns}}$. In the meantime, $v^{\mathcal{I}_{s}}$ is chosen to be the same as the first $k_s$ bits in $u^{\mathcal{I}_{ns}}$, denoted by $u^{\mathcal{I}_{ns}^{(1:k_s)}}$, as illustrated in Fig. \ref{fig:hybrid}. In this way, the values of pilots are not fixed, but dynamically change with the message bits. The connection between the non-systematic and systematic bits can be exploited by the decoder to perform channel estimation and also to improve error performance.

At the receiver side, the decoder recovers the message in a two-phase manner. In the first phase, it uses a CA-SCL decoder to find the most probable $\hat{u}^{\mathcal{I}_{ns}^{(1:k_s)}}$ with only the CDI of the channel. Note that unlike conventional CA-SCL decoder, this decoding process terminates as soon as the $k_s$-th non-frozen bit has been decoded, as we are only interested in $u^{\mathcal{I}_{ns}^{(1:k_s)}}$ in this phase. Since $v^{\mathcal{I}_{s}}=u^{\mathcal{I}_{ns}^{(1:k_s)}}$, $\hat{u}^{\mathcal{I}_{ns}^{(1:k_s)}}$ can be used for channel estimation. This phase is referred to as the \textit{channel estimation phase} in this paper. To achieve lower decoding error rate in this phase, $u^{\mathcal{I}_{ns}^{(1:k_s)}}$ can consist of $k_s-m_e$ ($m_e < k_s$) message bits and $m_e$ CRC bits of these message bits. In the second phase, the decoder recovers the rest message bits with the estimated channel gains and the knowledge of $\hat{u}^{\mathcal{I}_{ns}^{(1:k_s)}}$. We refer to this phase as the \textit{message decoding phase}. 

Besides playing the role of pilots for channel estimation, the systematic bits in our proposed scheme can also serve as check bits in the message decoding phase in addition to the CRC bits. For a decoding candidate $\hat{\mathbf{u}}$, besides checking its validity using CRC, we can also calculate $\hat{\mathbf{v}}=\hat{\mathbf{u}}\mathbf{G}_N$ and check whether the corresponding $\hat{v}^{\mathcal{I}_{s}}$ coincides with the decoded pilots $\hat{u}^{\mathcal{I}_{ns}^{(1:k_s)}}$ in the channel estimation phase. If not, this candidate is very likely to be wrong provided that  $\hat{u}^{\mathcal{I}_{ns}^{(1:k_s)}}$ is correct.

\subsection{Encoding}
Now we describe details of the encoding scheme. Suppose we use $m_e$ CRC bits for the channel estimation phase and $m_d$ CRC bits for the message decoding phase. The encoding procedures are summarized in Algorithm \ref{alg:1}.
\begin{algorithm}
	\caption{Encoding for Non-Coherent Communications}\label{alg:1}
	\begin{algorithmic}
		\Require the message vector $\mathbf{s}$
		\Ensure the codeword to be transmitted $\mathbf{x}$
		\State 1. The first $k_s-m_e$ message bits $s^{1:k_s-m_e}$ together with their length-$m_e$ CRC bits $c_e^{1:m_e}$ are assigned to $u^{\mathcal{I}_{ns}^{(1:k_s)}}$ and $v^{\mathcal{I}_{s}}$.
		\State 2. The rest message bits $s^{k_s-m_e+1:K}$ together with the length-$m_d$ CRC bits $c_s^{1:m_d}$ of $\mathbf{s}$ are assigned to the rest positions in $u^{\mathcal{I}_{ns}}$.
		\State 3. The hybrid polar encoder's output $\mathbf{v}$ is generated as described in Sec. \ref{Sec:HybScheme}.
		\State 4. The final codeword $\mathbf{x}$ is generated with an interleaver $\pi$ by $\mathbf{x}=\pi(\mathbf{v})$ as introduced previously.
	\end{algorithmic}
\end{algorithm}

\subsection{Decoding}
In the channel estimation phase, the decoder uses a CA-SCL decoder with list size $L_e$ to recover $u^{\mathcal{I}_{ns}^{(1:k_s)}}$. Since the correctness of the recovered pilots is vital for channel estimation and the subsequent message decoding phase, we may choose a relatively large $L_e$. As the number of pilots is much smaller than the length of the whole message, the increase of complexity induced by choosing a large $L_e$ can be controlled. The log-likelihood ratio (LLR) of $\mathbf{y}$ used in this decoder is calculated assuming that the channel gain is constant and equals $\mathbb{E}[h_i]$. Unlike conventional CA-SCL decoding, in this phase the SCL decoding procedure terminates when the $k_s$-th non-frozen bit has been decoded, and the most probable estimate $\hat{u}^{\mathcal{I}_{ns}^{(1:k_s)}}$ is selected with the help of CRC. Then $\hat{u}^{\mathcal{I}_{ns}^{(1:k_s)}}$ is used to estimate the channel gains of each block, denoted by $\hat{h}_i$, with some standard channel estimation method such as leat square (LS) or minimum mean square error (MMSE). 

In the message decoding phase, $u^{\mathcal{I}_{ns}^{(1:k_s)}}$ is treated as frozen bits with value $\hat{u}^{\mathcal{I}_{ns}^{(1:k_s)}}$. The log-likelihood ratios (LLRs) of $y_j$, $\pi^{-1}(j)\in \mathcal{I}_{s}$ (i.e., positions of pilots) are modified to infinity or minus infinity according to the corresponding value of $\hat{u}_i$ ($i\in \mathcal{I}_{ns}^{(1:k_s)}$). The LLRs of the rest $y_j$ are calculated according to the estimated channel gains. Then a conventional CA-SCL decoder with list size $L_d$ is used to recover the rest message bits.

The overall joint channel estimation and decoding procedures are summarized in Algorithm \ref{alg:2}.

\begin{algorithm}
	\caption{Joint Channel Estimation and Decoding}\label{alg:2}
	\begin{algorithmic}
		\Require the channel output $\mathbf{y}$
		\Ensure the estimate of the transmitted message $\hat{\mathbf{s}}$
		\State 1. Decode $\hat{u}^{\mathcal{I}_{ns}^{(1:k_s)}}$ with $\mathbf{y}$ and the CDI of the channel using a CA-SCL decoder of list size $L_e$.
		\State 2. Estimate the channel gains $\hat{h}_i$, $i=1,...,B$ with $\hat{u}^{\mathcal{I}_{ns}^{(1:k_s)}}$.
		\State 3. Update the LLRs of $y_j$ ($\pi^{-1}(j)\in \mathcal{I}_{s}$) by treating them as known bits, update the LLRs of the rest $y_j$ according to the estimated channel gains, set $u^{\mathcal{I}_{ns}^{(1:k_s)}}$ as frozen bits with value $\hat{u}^{\mathcal{I}_{ns}^{(1:k_s)}}$, and then perform CA-SCL decoding with list size $L_d$ to generate a list of candidates $\mathcal{L}=\{\hat{\mathbf{u}}_i\}$, $i=1,...,L_d$.
		\State 4. Find the most probable $\hat{\mathbf{u}}_i\in \mathcal{L}$, check whether it can pass CRC. If yes, go to 5, otherwise eliminate it from $\mathcal{L}$ and repeat 4.
		\State 5. Calculate $\hat{\mathbf{v}}_i=\hat{\mathbf{u}}_i\mathbf{G}_N$ and see if $\hat{v}_i^{\mathcal{I}_{s}}= \hat{u}^{\mathcal{I}_{ns}^{(1:k_s)}}$. If not, eliminate $\hat{\mathbf{u}}_i$ from $\mathcal{L}$ and go back to 4, otherwise use it to recover $\hat{\mathbf{s}}$ and terminate decoding.

	\end{algorithmic}
\end{algorithm}

\section{Simulation Results}
\label{Sec:Simulation}
In this section, we compare the performance of the proposed scheme with traditional polar codes. Since the frame error rate (FER) performance of systematic and non-systematic polar codes are the same, we use the performance of non-systematic polar codes as a benchmark. In our simulations, we assume $h_i$ follows the Rayleigh distribution with PDF
\begin{equation}
	f(h_i)=\frac{h_i}{\sigma_h^2}e^{-\frac{h_i^2}{2\sigma_h^2}},
\end{equation}
where $\sigma_h=1$. For traditional polar codes, we consider two cases, i.e., the instantaneous CSI is known by the receiver and only the CDI is known by the receiver. The decoding list size is 16 and the number of CRC bits is 16. For the proposed scheme, the decoding list size in the channel estimation phase and that in the message decoding phase are both 16. 16 CRC bits are used in total, with $m_e=12$ being allocated for the channel estimation phase and $m_d=4$ for the message decoding phase. To see the decoding error rate of the proposed scheme in the channel estimation phase, we also plot the error rates when decoding $u^{\mathcal{I}_{ns}^{(1:k_s)}}$ in the simulation results. 

\begin{figure}[tb]
	\centering
	\includegraphics[width=0.8\columnwidth]{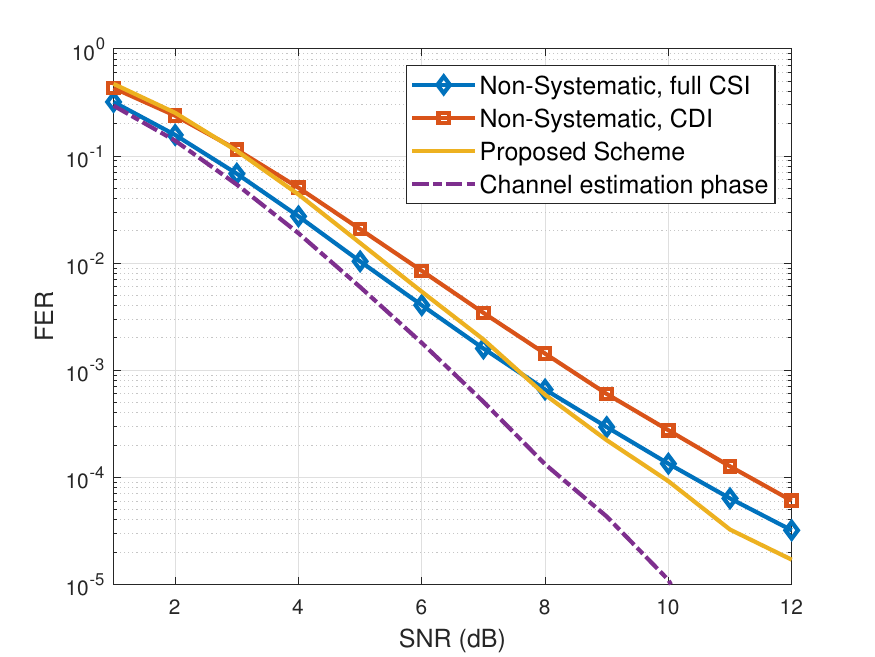} 
	\caption{Performance of the proposed scheme. $N=1024$, $R=1/2$, $T_c=64$, $B=16$.} 
	\label{fig:sim1}
\end{figure}

In the first example as shown in Fig. \ref{fig:sim1}, the channel coherent time is $T_c=64$ and the number of coherent blocks in a frame is $B=16$. Thus, the frame length is $N=1024$. The number of systematic bits in the proposed scheme is 48 (3 for each block). It can be seen that the performance of the proposed scheme is similar to that of non-systematic polar codes with only CDI at low SNRs, and gradually approaches and even outperforms that of non-systematic polar codes with perfect CSI as the SNR increases. 

It may look strange that the proposed scheme surpasses non-systematic polar codes with perfect CSI at high SNRs. This can be explained as follows. As can be seen from Fig. \ref{fig:sim1}, the decoding error probability of pilots (denoted by $P_e^p$ here) is much smaller than the overall FER. Suppose the pilots have been decoded correctly and the estimated channel gains are almost perfect (which is reasonable at high SNRs). Denote the error probability of the message decoding phase by $P_e^m$. It can be expected that $P_e^m$ is smaller than the FER of the non-systematic polar code with perfect CSI (denoted by $P_e^{ns}$), since the two codes have the same rate but LLRs at the pilot positions have been refined in the proposed scheme. The FER of the proposed scheme can be upper bounded by 
\begin{align}
	P_e^h\leq P_e^p+(1-P_e^p)P_e^m.
\end{align}
When $P_e^p\ll P_e^m$, we have $P_e^h\approx P_e^m<P_e^{ns}$. This explains why the proposed scheme is better than non-systematic polar coding with perfect CSI at high SNRs. 

We can also look at this problem from the viewpoint of concatenated codes. The proposed scheme can be seen as a concatenation of traditional polar code with a repetition code. Just like the concatenation with a CRC code can greatly improve the performance of polar codes under SCL decoding, the concatenation with a repetition code improves the performance of polar codes over non-coherent block fading channels.

\begin{figure}[tb]
	\centering
	\includegraphics[width=0.8\columnwidth]{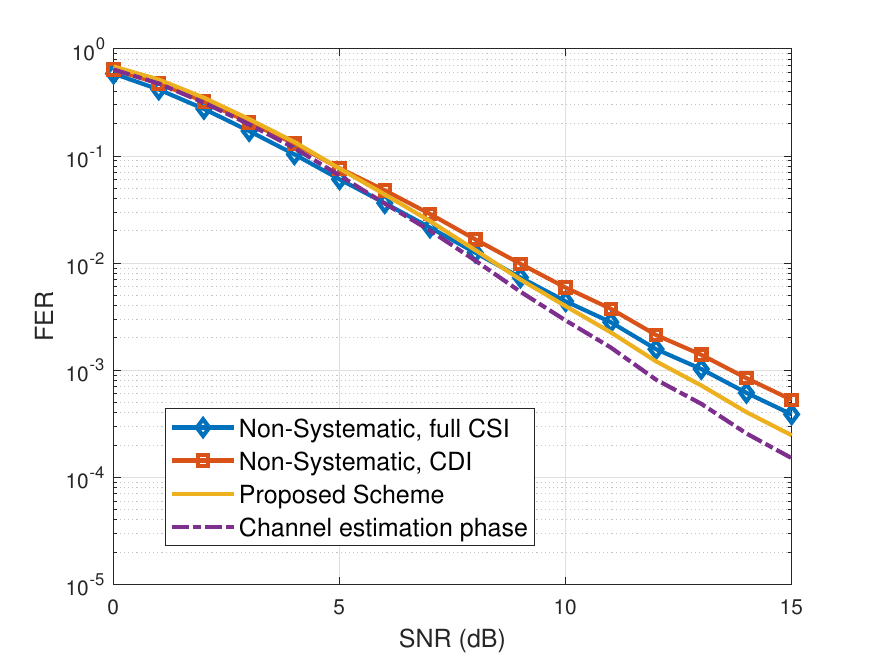} 
	\caption{Performance of the proposed scheme. $N=1024$, $R=1/2$, $T_c=128$, $B=8$.} 
	\label{fig:sim2}
\end{figure}

In the second example shown in Fig. \ref{fig:sim2}, the channel coherent time is $T_c=128$ and the number of coherent blocks in a frame is $B=8$. Thus, the frame length is also $N=1024$. The number of systematic bits in the proposed scheme is 48 (6 for each block). A similar result can be observed in this example.

\section{Conclusion}
\label{Sec:Conc}
In this paper we have proposed a polar coding scheme for non-coherent communications based on a hybrid encoding design that combines systematic and non-systematic polar coding. Connections between the non-systematic and systematic bits are exploited to perform joint channel estimation and decoding. Simulations have shown that the proposed scheme can even achieve better performance than traditional non-systematic polar codes with perfect CSI over some block fading channels at high SNRs. 

Although we have only considered BPSK modulation in this paper for the brevity of our description, the idea can be readily extended to higher order modulation and/or MIMO cases. We will investigate these scenarios in the future journal version of this work.

\bibliographystyle{IEEEtran}
\bibliography{HybPolar}

\end{document}